\documentclass[submit]{smj}

\usepackage{amssymb}
\usepackage{amssymb}
\usepackage{algorithmicx}
\usepackage{algcompatible}
\usepackage{algorithm}

\usepackage{lscape}
\usepackage{mathtools}
\usepackage{graphicx}
\usepackage{subfigure}
\usepackage{amsmath,amsthm,bm,bbm}

\usepackage[english]{babel}
\usepackage[shortlabels]{enumitem}

\newcommand{\prob}{{\mathbb{P}}}

\newcommand{\iidsim}{{\overset{\mathrm{i.i.d.}}{\sim}}}
\newcommand{\transpose}{^{\mathrm{T}}}

\newcommand{\calL}{{\mathcal{L}}}

\newcommand{\calN}{{\mathcal{N}}}

\newcommand{\bM}{{\mathbf{M}}}

\newcommand{\bY}{{\mathbf{Y}}}

\newcommand{\bx}{{\mathbf{x}}}
\newcommand{\bX}{{\mathbf{X}}}

\newcommand{\by}{{\mathbf{y}}}
\newcommand{\bz}{{\mathbf{z}}}

\newcommand{\bb}{{\mathbf{b}}}

\newcommand{\bbeta}{{\bm{\beta}}}

\newcommand{\bOmega}{{\bm{\Omega}}}

\newcommand{\bSigma}{{\bm{\Sigma}}}

\newcommand{\eye}{{\mathbf{I}}}

\newcommand{\btheta}{{\bm{\theta}}}
\newcommand{\bmu}{{\bm{\mu}}}

\newcommand{\zero}{{\bm{0}}}
\newcommand{\eps}{\epsilon}

\Author{Qiushuang Li\Affil{1}\and
        Recai M. Yucel\Affil{1}, 
        
}
\AuthorRunning{Qiushuang Li \and Recai M. Yucel}

\Affiliations{

\item Department of Epidemiology and Biostatistics, 
      College of Public Health, 
      Temple University,
      Philadelphia,
      United States
}   

\CorrAddress{Recai M. Yucel, 
             Department of Epidemiology and Biostatistics, 
      College of Public Health, 
      Temple University,
      1101 W. Montgomery Ave, 
      Philadelphia,
      United States}
\CorrEmail{recai.yucel@temple.edu}
\CorrPhone{(+1)\;215\;204\;6240}

\Title{Sequential Hierarchical Regression Imputation with Variable Selection Routines}
\TitleRunning{Regression Imputation with Variable Selection}

\Abstract{
We aim to incorporate variable selection routines into variable-by-variable (or sequential) imputation in clustered data to achieve computational improvement in applications with large-scale health data. Specifically, we utilize variable selection routines using spike-and-slab priors within the Bayesian variable selection routine. The choice of these priors allows us to ``force'' variables of importance (e.g., design variables or variables known to play a role in the missingness mechanism) into the imputation models based on a class of mixed-effects models. Our ultimate goal is to improve computational speed by removing unnecessary variables.  We employ Markov chain Monte Carlo techniques to sample from the implied posterior distributions for model unknowns as well as missing data.  We assess the performance of our proposed methodology via simulation studies. Our results show that our proposed algorithms lead to satisfactory estimates and, in some instances, outperform some of the existing methods that are available to practitioners. We illustrate our methods using a national survey of children's health.
}


\begin{document}

\maketitle

\newpage
\section{Introduction\label{sec:Introduction}}
Statistically sound methods for missing data have been of interest in many problems in a wide variety of disciplines. In statistical analysis of high-dimensional data, it is common to encounter large covariance matrix estimation problems for various purposes, such as dimension reduction, graphical modeling of conditional independence of random variables via structured learning, and image processing.  These analytical aims are typically complicated by arbitrary missing values \citep{lounici2014high}. 
Survey data are also subject to missing data but are often more complicated due to skip patterns or bounds.  In other fields, missing data can occur in computer experiments as well as biomedical applications due to equipment limitations \cite{bayarri2007framework}. Essentially,  missing data are norm rather than exceptional in a broad range of fields, and sensible inferences require thoughtful considerations to counter the potentially adverse impact of missing data.

The method of imputation has long been in practice  when dealing with missing data.  Regardless of how methodologically advanced it can be,  single imputation methods are known to be problematic as they can lead to inaccurate estimation of statistical uncertainty and potential estimation bias. The idea of multiple imputation (MI), which was first introduced by \cite{rubinmultiple,10.1093/biomet/63.3.581}, has become a standard method to account for uncertainty due to missing data.   MI proposes to sample from a plausible predictive distribution of missing data so that  uncertainty due to missingness is accounted for in the analyses. The statistical analysis proceeds by treating each set of the imputed data as a set of complete data, followed by a combined analysis using Rubin's method \citep{rubinmultiple,10.1093/biomet/63.3.581}. More specifically, the MI is built upon a complete probabilistic model for the complete data,  from which a simulation-based approach is implemented to perform multiple imputations for the missing portion. 

Statistical computation underlying MI is typically based on either joint  or variable-by-variable (or sequential) imputation models 
\cite{schafer1997analysis,gelman2004parameterization,heckerman2000dependency,kennickell1991imputation,raghunathan2001multivariate,buuren2010mice,https://doi.org/10.1111/1467-9574.00217,yucel2017sequential,liu2014stationary}. In either framework, the most common strategy for MI is based on Bayesian modeling by drawing MI samples from the posterior predictive distributions of the missing data. For example, in the variable-by-variable imputation framework, one begins first by specifying the conditional distribution of the complete data given the unknown parameters, often referred to as the complete-data likelihood, and then the distribution for the unknown parameters, referred to as prior distributions. $[$This is based on the specification of the full conditional distributions of each variable given the remaining variables and iteratively drawing missing values from these full conditionals, a procedure that quite resembles the classical Gibbs sampler.$]$ This is followed by a posterior computation via a Markov chain Monte Carlo sampler that draws random samples from the posterior distribution of the unknown parameters as well as the missing portion of the data given the observed portion of the data. Then, each random sample drawn from the posterior predictive distribution of the missing data forms an imputed version of the missing values.

In our work, we also consider the variable selection problem.  The variable selection problem arises in regression models when the number of available predictors or covariates to users exceeds the number of true active predictors, and one aims to recover the correct set of active predictors. There has been vast literature on developing frequentist methods for variable selection. Classical criterion-based approaches include generalized cross-validation (GCV) and the Bayesian information criterion (BIC). These methods become computationally expensive when the number of candidate predictors becomes large as they require exhaustive searches of all possible sub-models, the number of which grows exponentially with the number of predictors. The last decade has also witnessed the progress of penalized-based approaches for variable selection \citep{bickel2006regularization}, including the LASSO, Smoothly Clipped Absolute Deviation (SCAD) penalty \citep{zou2006adaptive}, and Adaptive LASSO (ALASSO) \citep{zou2006adaptive}. These methods translate the problem of variable selection into convex programming problems, and there have been relatively mature algorithms for solving these mathematical optimization problems, greatly facilitating the use of penalized likelihood methods. 

Significant progress has also been made in developing Bayesian methods for variable selection. The most widely adopted method is via the spike-and-slab prior distribution \citep{castillo2012needles,castillo2015bayesian}. In particular, \cite{castillo2015bayesian} extensively studied the theoretical properties of the Bayesian linear regression model with fixed effects using the spike-and-slab prior distribution. Other forms of the variable selection prior include the Bayesian LASSO \citep{doi:10.1198/016214508000000337}, the horseshoe prior \cite{carvalho2010horseshoe}, the Dirichlet-Laplace prior \citep{doi:10.1080/01621459.2014.960967}, and the spike-and-slab LASSO prior \citep{rovckova2018bayesian,rovckova2018spike}. This body of literature, however, focuses on sparsity recovery and parameter estimation in regression models and do not consider missing data scenario as well as MI, which is the focus of this work. 

There has also been some progress in incorporating variable selection methods in the context of missing data analysis. One strategy is to focus on parameter estimation and inference without MI using the incomplete-data likelihood, and this line of work includes \cite{https://doi.org/10.1111/j.1541-0420.2009.01274.x,garcia2010variable}. The challenge of these likelihood-based methods is that they require the computation of the likelihood function of incomplete data when one is faced with missing responses and/or predictors. Such incomplete-data likelihoods are typically intractable to compute and involve high-dimensional integrals \citep{garcia2010variable}. These methods rely on EM algorithms and are not easily extended to broader contexts. Another strategy is to tackle parameter estimation and MI simultaneously. The problem of variable selection across MI has been a longstanding challenge because the variable selection outcomes may not coincide with each other across different MI copies if it is performed respectively for each MI copy. Combining variable selection results from different MI copies is challenging. \cite{Heymans2007,https://doi.org/10.1002/sim.3177,lachenbruch2011variable} proposed to include variables that are selected at least $\pi M$ amount of times across $M$ MI copies, where $\pi$ is a selection threshold between $0$ and $1$. Other frequentist approaches based on bootstrap and penalized methods include \cite{https://doi.org/10.1002/sim.5783,https://doi.org/10.1002/sim.3177,liu2019imputation,10.1214/15-AOAS899,10.1093/biostatistics/kxv003}. For a review, see \cite{zhao2017variable}. The aforementioned literature is largely based on frequentist methods, and there is comparatively narrower development in combining variable selection and MI in a coherent Bayesian framework. In this regard, our work is similar to \cite{https://doi.org/10.1111/j.1541-0420.2005.00317.x} in the sense of simultaneous variable selection and MI using Bayesian methods and MCMC. Still, our work extends \cite{https://doi.org/10.1111/j.1541-0420.2005.00317.x} by considering clustered data and mixed-effect models as well as generalized linear models, and it can be put in the context of variable-by-variable imputation framework. 

Specifically, we consider Bayesian methods for variable selection and deal with the problem of missing data using generalized linear mixed-effects models as the basis for drawing MIs.
The proposed method can simultaneously perform variable selection and multiple imputations of missing responses for continuous and binary responses via mixed-effects models. For computation, the key technical challenge is that the full posterior distributions of some of the parameters are not in closed form,  making the Markov chain Monte Carlo sampler terribly cumbersome to implement. To this end, we consider specialized computational techniques by introducing the P\'olya-Gamma auxiliary variables originally due to \citep{doi:10.1080/01621459.2013.829001} to bypass this problem. The detailed methods are discussed in Section \ref{sec:logistic_mixed_effects_regression_models}.

To draw the imputations, we adopt the idea of variable-by-variable imputation routines (such as \cite{yucel2017sequential}) and draw each of the missing variables through a conditional regression imputation model sequentially. Roughly speaking, variable-by-variable imputation routines refer to a collection of multiple imputation methods where each variable is modeled conditionally given the remaining variables in an imputation model, and the missing data of each variable is imputed in turn conditionally. In many settings, especially those underlie surveys, these methods can be great alternatives to those relying on the joint models. Additionally, the advantage of this variable-by-variable MI strategy is that it significantly reduces the computational complexity and burden for high-dimensional data \citep{yucel2017sequential}.
The formal description of this strategy will be introduced in Section \ref{sec:linear_mixed_effects_regression_models}.

The rest of this paper is organized as follows. Section \ref{sec:linear_mixed_effects_regression_models} 
discusses the computational algorithm based on the Gibbs sampler for 
 linear mixed-effects regression models with spike-and-slab priors for variable selection when the response variables are continuous. We then extend these methods to binary response variables using the logistic mixed-effects model in Section \ref{sec:logistic_mixed_effects_regression_models}, introduce the P\'olya-Gamma random variables, leverage them for the parameter expansion for data augmentation (PX-DA) \cite{liu1999parameter}, and develop a closed-form Gibbs sampler. These two Gibbs samplers allow simultaneous inference of the parameters while drawing samples for the missing responses. The advantage of the proposed approach is illustrated using extensive simulated examples in Section \ref{sec:simulated_examples}. Section 5 presents an application of our algorithms to the National Survey of Children’s Health (NSCH).  We conclude with discussion in Section \ref{sec:discussion}.

\section{Models}
\label{sec:models}

Throughout this work, we let $\by$ denote variables subject to missingness, and  $\bx_{ij}$ will denote the covariate vector for the $j$th subject in the $i$th cluster that is fully observed or imputed in a similar manner presented here. Depending on the measurement scales, we will utilize a linear mixed-effect model for continuous variables or a logistic mixed-effect model for binary variables. The missingness mechanism is assumed to be missing at random (MAR) throughout this work. Formally, we use $\mathbf{R}_\by = [R_{y_{ij}}]_{i,j}$ to denote the binary random variables encoding whether the elements of $\mathbf{R}_\by$ are missing or observed: $R_{y_{ij}} = 1$ if $y_{ij}$ is missing and $R_{y_{ij}} = 0$ if $y_{ij}$ is observed. Then MAR states that the distribution of $\mathbf{R}_\bY$ only depends on the observed data $\mathbb{X}$ but not on the missing data, $\bY$, itself. 

\subsection{Model for continuous outcomes}
\label{sec:linear_mixed_effects_regression_models}

Similar to \cite{yucel2017sequential}, we consider a linear mixed-effects model with random intercept only for continuous response variable $y_{ij}$:
\begin{align}
\label{eqn:LME_model}
y_{ij} = \bx_{ij}\transpose\bbeta + b_i + \eps_{ij},\quad i = 1,\ldots,m,\quad j = 1,\ldots,n_i,
\end{align}
where $\bbeta\in\mathbb{R}^p$ is the fixed-effect regression coefficient, $b_1,\ldots,b_m\iidsim\mathrm{N}(0,\sigma_b^2)$ are the random effects, and $\eps_{ij},\iidsim\mathrm{N}(0, \sigma_e^2)$ are random errors, $i=1,2,\ldots,m, j=1,2,\ldots,n_i$. Extension to cases where additional random-effect covariates are available is straightforward. Subscripts $i$ and $j$ denote the cluster and observation of that cluster, respectively.   The responses $y_{ij}$'s are either observed or missing and the missing portion will be imputed using a random draw from its underlying posterior distribution via a Gibbs sampler, as detailed below. Finally, $\bx_{ij}\in\mathbb{R}^p$'s are the individual-level covariates that can also be either observed or missing, and the missing values are imputed via the last cycle of the SHRIMP strategy, as is suggested in \cite{yucel2017sequential}.

We develop a Gibbs sampler to draw independent samples from the joint posterior distribution of $(\bbeta, b_1,\ldots,b_m, \sigma_b, \sigma_e)$, as well as to draw samples of the missing data $(y_{\mathrm{mis}})$.
To select the variables among $x_{ij1},\ldots,x_{ijp}$, we use a spike-and-slab prior distribution, which has been widely applied in Bayesian variable selection methods \citep{doi:10.1080/01621459.1988.10478694,doi:10.1080/01621459.1993.10476353,doi:10.1080/01621459.1996.10476989,geweke1996variable,10.2307/25053023}. Specifically, it is imposed on the fixed-effects coefficient $\beta_k$. In missing data applications where the $k^{th}$ variable is supposed to be clearly not important, then we could assign the following spike-and-slab prior to $\beta_k$:
\begin{align}\label{eqn:spike_and_slab_prior}
\beta_k\mid w, \mu_0,\sigma_0&\left\{
\begin{array}{ll}
= 0,&\quad\text{with probability }(1 - w),\\
\sim\mathrm{N}(\mu_0, \sigma_0^2),&\quad\text{with probability }w,
\end{array}
\right.
\end{align}
where $w$ is the prior probability that the $k$th variable $x_{ijk}$ is selected, and with probability $(1 - w)$, $\beta_k$ is set to $0$ so that under the prior distribution, the $k$th variable is not selected. 
The spike-and-slab prior distribution \eqref{eqn:spike_and_slab_prior} can be equivalently written as
\[
(\beta_k\mid w, \mu_0,\sigma_0)\sim (1 - w)\delta_0 + w\mathrm{N}(\mu_0, \sigma_0^2),
\]
where $\delta_0$ is a point mass at $0$. Otherwise, $\beta_k$ is assigned a normal prior if there is a sure certainty of selection:
\[
(\beta_k\mid w, \mu_0,\sigma_0)\sim \mathrm{N}(\mu_0, \sigma_0^2).
\]
To reduce the effect of hyperparameters and enhance the robustness of the entire Bayesian model, we further assume that the hyperparameters have the following hyperprior distributions: $w\sim\mathrm{Beta}(a_w, b_w)$, $\mu_0\sim\mathrm{N}(0, 1)$, and $\sigma^2_0\sim\mbox{Inverse-Gamma}(1, 1)$. For the rest of the parameters $(\sigma_b^2, \sigma_e^2)$, we assume the inverse-$\chi^2$ distribution for the sake of conjugacy:  $\sigma_b^2\sim\chi_{\nu_b}^{-2}$ and $\sigma_e^2\sim\chi_{\nu_e}^{-2}$. 

We provide the detailed full conditional distributions that underlie the Gibbs sampler in  \ref{sec:gibbs_sampler_for_section_sec:linear_mixed_effects_regression_models}. Here, we focus on the conditional distribution of the linear coefficients $\beta_k$, $k = 1,2,\ldots,p$.
Denote by the parameters $\btheta_{-k}$ the set of all parameters except $\beta_k$: $\btheta_{-k} = (\bbeta_{-k}, \sigma_b, \sigma_e)$, where $\bbeta_{-k} = \{\beta_1,\ldots,\beta_p\}\backslash\{\beta_k\}$, and the random effects $\bb = [b_1,\ldots,b_m]\transpose$. Then the full conditional distribution of $\beta_k$ for $k = 1,2,\ldots,p$ is given by
\begin{align}
\label{eqn:Gibbbs_beta}
&(\beta_k\mid\bX, \btheta_{-k}, w, \mu_0, \sigma_0)\sim
\left\{
\begin{array}{ll}
 w^*_1\delta_0 + w^*_2\mathrm{N}(\widehat\mu, \widehat V),&\text{if  the }k\text{th variable is undetermined},\\
 \mathrm{N}(\widehat\mu, \widehat V),&\text{if  the }k\text{th variable is forced to be selected},
\end{array}
\right.
\end{align}
where $\bX$ denotes the full set of covariates $\bX = \{\bx_{i1},\ldots,\bx_{in_i}\}_{i = 1}^m$, and the formulas for $w_1^*,w_2^*,\widehat{V},\widehat{\mu}$ are listed below:
\begin{align*}
&w_1^*\propto (1 - w)
\calN\left(0\mathrel{\Big|}\frac{\sum_{i,j}x_{ijk}(y_{ij} - \sum_{\ell\neq k}x_{ij\ell}\beta_\ell - b_i)}{\sum_{i,j}x_{ijk}^2}, \frac{\sigma_e^2}{\sum_{i,j} x_{ijk}^2}\right),\\
&w_2^*\propto w
\calN\left(\mu_0\mathrel{\Big|}\frac{\sum_{i,j}x_{ijk}(y_{ij} - \sum_{\ell\neq k}x_{ij\ell}\beta_\ell - b_i)}{\sum_{i,j}x_{ijk}^2}, \sigma_0^2 + \frac{\sigma_e^2}{\sum_{i,j} x_{ijk}^2}\right),\\
&\widehat V = \left(\frac{1}{\sigma_e^2}\sum_{i = 1}^m\sum_{j = 1}^{n_i}x_{ijk}^2 + \frac{1}{\sigma_0^2}\right)^{-1},\;
\widehat\mu = \widehat V\left[\frac{\mu_0}{\sigma_0^2} + \frac{1}{\sigma_e^2}\sum_{i = 1}^m\sum_{j = 1}^{n_i}x_{ijk}\left(y_{ij} - \sum_{\ell\neq k}x_{ij\ell}\beta_\ell - b_i\right)\right],
\end{align*}
where $\calN(x\mid\mu, v^2):=(2\pi v^2)^{-1/2}e^{-(x - \mu)^2/(2v^2)}$ denotes the probability density function of $\mathrm{N}(\mu, v^2)$ evaluated at $x$. The derivation of the rest of the full conditional distributions is routine and is provided in \ref{sec:gibbs_sampler_for_section_sec:linear_mixed_effects_regression_models}.
We also emphasize that \eqref{eqn:Gibbbs_beta} presents the nature of variable selection inside a single cycle of the Gibbs sampler: with probability $w_1^*$, we set $\beta_k = 0$, suggesting that currently, the $k$th variable is not selected, and with probability $w_2^*$, we draw $\beta_k$ from a normal distribution, indicating that $\beta_k\neq 0$, and therefore, the $k$th variable needs to be selected. 

We also remark that because we assign the spike-and-slab prior for the regression coefficients $\beta_1,\ldots,\beta_p$ independently rather than assigning a joint multivariate normal distribution, the joint posterior distribution is no longer a multivariate normal but a much more complicated distribution. This is quite different from the usual multiple imputation approaches where the regression coefficient vector is drawn jointly from a multivariate normal. To tackle the computation of such a generally intractable posterior distribution, we propose to draw from the full conditional distribution of each $\beta_k$ within a Gibbs sampler, as detailed above. 

\subsection{Model for binary outcomes
} 
\label{sec:logistic_mixed_effects_regression_models}

We use the following conventional logistic mixed-effects regression as the basis to draw missing values in binary variables:
\[
\prob\left(y_{ij} = 1\mid\bx_{ij}, b_i, \bbeta\right) = \frac{1}{1 + \exp(-\bx_{ij}\transpose\bbeta - b_i)},
\]
where $\bbeta$ are the fixed-effects coefficients for covariates $x_i$, and $b_1,\ldots,b_m\iidsim\mathrm{N}(0, \sigma_b^2)$ are the random effects. To develop a closed-form Gibbs sampler to draw from the posterior distribution of unknowns under this model,  we adopt a similar strategy suggested by \cite{doi:10.1080/01621459.2013.829001}. They suggest introducing a collection of auxiliary variables following the \emph{P\'olya-Gamma} distribution, such that the full conditional distributions of all parameters are available in closed form. We first present the definition of the P\'olya-Gamma distribution (see Definition 1 in \citealp{doi:10.1080/01621459.2013.829001}): A random variable $X$ is said to follow a P\'olya-Gamma distribution with parameters $b > 0$ and $c\in\mathbb{R}$, denoted by $X\sim\mathrm{PG}(b, c)$, if there exists a sequence of independent Gamma random variables $(g_k)_{k = 1}^\infty\iidsim \mathrm{Gamma}(b, 1)$, such that
\[
X = \frac{1}{2\pi^2}\sum_{k = 1}^\infty\frac{g_k}{(k - 1/2)^2 + c^2/(4\pi^2)}.
\]
We utilize the key result of the P\'olya-Gamma distribution (see Theorem 1 in \citep{doi:10.1080/01621459.2013.829001}), which states that if $p(\omega)$ is the density function of $\omega\sim \mathrm{PG}(b, 0)$, $b > 0$, then the following integral identity holds for all $a\in\mathbb{R}$:
\[
\frac{[\exp(\psi)]^a}{[1 + \exp(\psi)]^b} = 2^{-b}\exp\left[\left(a - \frac{b}{2}\right)\psi\right]\int_0^\infty \exp\left( - \frac{1}{2}\omega\psi^2\right)p(\omega)\mathrm{d}\omega.
\]
Moreover, in the above case, 
$p(\omega\mid\psi) = \exp\left( - \omega\psi^2/2\right)p(\omega)/{\int_0^\infty \exp\left( - \omega\psi^2/2\right)p(\omega)\mathrm{d}\omega}$
is the density function of $\omega\sim\mathrm{PG}(b, \psi)$.

We let the following prior distributions reflect the appropriate prior knowledge on the fixed-effects coefficients $\beta_1,\ldots,\beta_p$. Similar to the continuous response modelling, we assign a spike-and-slab prior \eqref{eqn:spike_and_slab_prior} to $\beta_1,\ldots,\beta_p$ as follows
\begin{align}
\begin{aligned}
&(\beta_k\mid w,\mu_0,\sigma_0^2)\sim (1 - w)\delta_0 + w\mathrm{N}(\mu_0, \sigma_0^2),\quad \text{if the }k\text{th variable is undetermined},\\
&(\beta_k\mid w,\mu_0,\sigma_0^2)\sim \mathrm{N}(\mu_0, \sigma_0^2),\quad \text{if the }k\text{th variable is forced to be selected},\\
&w\sim\mathrm{Beta}(a_w, b_w),\quad \mu_0\sim\mathrm{N}(0, 1),\quad \sigma^2\sim\mbox{Inverse-Gamma}(1, 1).
\end{aligned}
\end{align}
The prior distribution on $\sigma_b$ is same as  Section \ref{sec:linear_mixed_effects_regression_models}: $\sigma_b^2\sim\chi_{\nu_b}^{-2}$. 

We now elaborate on the full conditional distributions of the linear coefficients $\beta_k$, $k = 1,2,\ldots,p$. The rest of the full conditional distributions to implement the Gibbs sampler for drawing independent draws from the joint posterior distribution of $(\bbeta, b_1,\ldots,b_m)$, together with the samples of the missing data $(y_{\mathrm{mis}})$, are provided in \ref{sec:gibbs_sampler_for_section_sec:logistic_mixed_effects_regression_models}. Following the derivation in \cite{doi:10.1080/01621459.2013.829001}, we derive the likelihood function of $\eta_{ij}:= \bx_{ij}\transpose \bbeta + b_i$:
\begin{align*}
\calL(\eta_{ij}\mid y_{ij})
& \propto \exp\left[\left(y_{ij} - \frac{1}{2}\right)\eta_{ij}\right]\int_0^\infty \exp\left(-\frac{\omega_{ij}\eta_{ij}^2}{2}\right)p(\omega_{ij}\mid 1,0)\mathrm{d}\omega_{ij},
\end{align*}
where $p(\omega_{ij}\mid1,0)$ is the density of an auxiliary P\'olya-Gamma random variable $\omega_{ij}\sim\mathrm{PG}(1, 0)$. 
The idea of introducing the auxiliary variables $\omega_{ij}$'s is such that after marginalizing them out, the joint distribution of the rest variables is left invariant. We derive the likelihood of $\bbeta$ for all $mn$ data points after introducing $\bOmega = \{\omega_{i1},\ldots,\omega_{in_i}\}_{i = 1}^m$:
\begin{align*}
&\calL(\bbeta\mid \bX, \bY, \bOmega, b_1,\ldots,b_m,\sigma^2)
\propto \exp\left\{
-\frac{1}{2}(\bz - \bX\bbeta)\transpose\bSigma^{-1}(\bz - \bX\bbeta)\right\},
\end{align*}
where $z_{ij} = {(y_{ij} - 1/2)}/{\omega_{ij}} - b_i$,
\begin{align*}
\bz & = \left[z_{11},\ldots,z_{1n_1},z_{21},\ldots,z_{2n_2},\ldots,z_{m1},\ldots,z_{mn_m}\right]\transpose\in\mathbb{R}^{\sum_in_i},\\
\bX & = [\bx_{11},\ldots,\bx_{1n_1},\bx_{21},\ldots,\bx_{2n_2},\ldots,\bx_{m1},\ldots,\bx_{mn_m}]\transpose\in\mathbb{R}^{(\sum_in_i)\times p},\\
\bSigma^{-1} &= \mathrm{diag}(\omega_{11},\ldots,\omega_{1n_1},\omega_{21},\ldots,\omega_{2n_2},\ldots,\omega_{m1},\ldots,\omega_{mn_m})\in\mathbb{R}^{(\sum_in_i)\times (\sum_in_i)}.
\end{align*}
We then obtain the following closed-form full conditional distribution of $\beta_k$, $k = 1,2,\ldots,p$:
\begin{align}
\label{eqn:binary_Gibbs_beta}
&(\beta_k\mid\bX, \bY, \bOmega, \bbeta_{-k}, \bb,\sigma_b, w, \mu_0, \sigma_0)\\
&\quad\sim
\left\{
\begin{array}{ll}
 w^*_1\delta_0 + w^*_2\mathrm{N}(\widehat\mu, \widehat V),&\text{if  the }k\text{th variable is undetermined},\\
 \mathrm{N}(\widehat\mu, \widehat V),&\text{if  the }k\text{th variable is forced to be selected},
 \end{array}
\right.
\end{align}
where $\bX$ denotes the full set of covariates $\bX = \{\bx_{i1},\ldots,\bx_{in_i}\}_{i = 1}^m$, and
\begin{align*}
&w_1^*\propto (1 - w)
\calN\left(0\mathrel{\Bigg|}\frac{\sum_{i,j}\omega_{ij}x_{ijk}(z_{ij} - \sum_{\ell\neq k}x_{ij\ell}\beta_\ell)}{\sum_{i,j}\omega_{ij}x_{ijk}^2}, \frac{1}{\sum_{i,j} \omega_{ij}x_{ijk}^2}\right),\\
&w_2^*\propto w
\calN\left(\mu_0\mathrel{\Bigg|}\frac{\sum_{i,j}\omega_{ij}x_{ijk}(z_{ij} - \sum_{\ell\neq k}x_{ij\ell}\beta_\ell)}{\sum_{i,j}\omega_{ij}x_{ijk}^2}, \sigma_0^2 + \frac{1}{\sum_{i,j} \omega_{ij}x_{ijk}^2}\right),\\
&\widehat V = \left(\sum_{i = 1}^m\sum_{j = 1}^{n_i}\omega_{ij}x_{ijk}^2 + \frac{1}{\sigma_0^2}\right)^{-1},\;
\widehat\mu = \widehat V\left[\frac{\mu_0}{\sigma_0^2} + \sum_{i = 1}^m\sum_{j = 1}^{n_i}\omega_{ij}x_{ijk}\left(z_{ij} - \sum_{\ell\neq k}x_{ij\ell}\beta_\ell\right)\right].
\end{align*}
The full conditional distribution of the auxiliary variables $\bOmega = \{\omega_{i1},\ldots,\omega_{in_i}\}_{i = 1}^m$ can be derived similarly as that in \cite{doi:10.1080/01621459.2013.829001}:
\begin{align}
\label{eqn:binary_Gibbs_omega}
(\omega_{ij}\mid\bbeta, b_1,\ldots,b_m)\sim\mathrm{PG}(1, \bx_{ij}\transpose\bbeta + b_i),
\end{align}
and sampling a random variable following a P\'olya-Gamma distribution can be implemented using the algorithm described in Section 4 in \cite{doi:10.1080/01621459.2013.829001}. 
The derivation of the rest of the full conditional distributions is similar to those in Section \ref{sec:linear_mixed_effects_regression_models}, and we leave them in  \ref{sec:gibbs_sampler_for_section_sec:logistic_mixed_effects_regression_models}.
Finally, for each missing $y_{ij}\in(y_{\mathrm{mis}})$, one can draw it from the following conditional distribution in a single cycle of the Gibbs sampler:
\[
(y_{ij}\mid\bX, \bbeta, b_1,\ldots,b_m)\sim\mathrm{Bernoulli}\left(\frac{1}{1 + \exp(-\bx_{ij}\transpose\bbeta - b_i)}\right).
\]
Similar to our algorithm of Gibbs sampler in Section \ref{sec:linear_mixed_effects_regression_models}, the above procedure is performed to draw values for the binary variables.

\section{Efficient sequential hierarchical regression imputation}
\label{sec:SS_SHRIMP}

Sequential hierarchical regression imputation (SHRIMP) \cite{yucel2017sequential} has been a powerful tool for missing data, especially in survey settings where skip patterns, restriction and diverse measurement scales can be typical. Such scenarios can substantially complicate drawing MIs under joint models for the variables subject to missing data. 

This section incorporates the proposed modeling approach as well as the computational algorithms to SHRIMP.  Let $\bM_1,\ldots,\bM_m$ denote the data across the $m$ clusters, with each $\bM_i$ being an $n_i\times d$ matrix, where rows represent observations and columns represent variables. The first step of SHRIMP is to sort the variables according to their respective missing proportions. Formally, we order the variable indices $\{1,2,\ldots,d\}$ such that the sorted indices, say $\{k_1,\ldots,k_d\}$, satisfy
\[
\sum_{i = 1}^m\sum_{j = 1}^{n_i}\mathbbm{1}([\bM_i]_{jk_l} = \text{NA})\leq \sum_{i = 1}^m\sum_{j = 1}^{n_i}\mathbbm{1}([\bM_i]_{jk_{l + 1}} = \text{NA}),\quad l = 1,\ldots,d - 1,
\]
where $[\bM_i]_{jk}$ denotes the $(j, k)$th entry of $\bM_i$, i.e., the number of missing values of the $k_l$th variable is always no greater than that of the $k_{l + 1}$th variable.   

Next, according to this order, SHRIMP proceeds with the imputation process in a variable-by-variable fashion. Specifically, one cycle of the SHRIMP consists of the following operations. Assume that the previous cycle of SHRIMP has already produced imputed values for missing entries of $(\bM_i)_{i = 1}^m$. For each $k = k_1,k_2,\ldots,k_d$, let $y_{ij}^{(k)}$ be the $j^{th}$ observation of the $k^{th}$ variable in the $i^{th}$ cluster, and $\bx_{ij}^{(k)}$ be the remaining variables of the $j^{th}$ observation in the $i^{th}$ cluster. Let $\bY = \{y_{i1}^{(k)},\ldots,y_{in_i}^{(k)}\}_{i = 1}^m$, and $\bX = \{\bx_{i1}^{(k)},\ldots,\bx_{in_i}^{(k)}\}_{i = 1}^m$. Here, the missing values of $\bY$ remain, but the missing values of $\bX$ are imputed with values generated from the previous cycle of SHRIMP. 
    \begin{itemize}
      \item If the $k$th variable is continuous, use the linear mixed-effect model in Section \ref{sec:linear_mixed_effects_regression_models} as the conditional imputation model. Run the Gibbs sampler in Section \ref{sec:linear_mixed_effects_regression_models} with the above $\bX$ and $\bY$. Return the imputed values generated from the Gibbs sampler to the original data in $(\bM_i)_{i = 1}^m$. 
      \item If the $k$th variable is binary, use the logistic mixed-effect model in Section \ref{sec:logistic_mixed_effects_regression_models} as the conditional imputation model. Run the Gibbs sampler in Section \ref{sec:logistic_mixed_effects_regression_models} with the above $\bX$ and $\bY$. Return the imputed values generated from the Gibbs sampler to the original data in $(\bM_i)_{i = 1}^m$. 
    \end{itemize}
By iterating the above cycles for a sufficiently large number of times within each step above to ensure that the underlying MCMC converges, we are able to obtain a sequence of samples of missing entries of $(\bM_i)_{i = 1}^m$ 
which are approximate draws from 
$P(\bM_{\mathrm{mis}}\mid \bM_{\mathrm{obs}})$ as the number of cycles goes to infinity, where $\bM_{\mathrm{obs}}$ and $\bM_{\mathrm{mis}}$ denote the observed and missing portion of $(\bM_i)_{i = 1}^m$. 
After the SHRIMP is completed, the final set of drawn values of all the missing entries forms one copy of the imputation. For the purpose of multiple imputations, one can repeat this procedure for $M$ times to obtain $M$ copies of the imputed data. 

\section{Simulation Study \label{sec:simulated_examples}}
The purpose of this study is to assess sampling properties of the underlying MI inference where imputed values are drawn using our proposed algorithm. We establish this goal by (a) repeatedly sampling data as described below; (b) imposing missing values under MAR mechanism; (c) drawing missing values under the spike-and-slab sequential hierarchical regression imputation (SS-SHRIMP) introduced in Section \ref{sec:SS_SHRIMP} to form the multiple imputations; and finally (d) fit a hypothetical analysis model and obtain MI estimates and assess the underlying criteria gauging their quality such as coverage rates, mean square error (MSE) as well as fraction of missing information (FMI).

\textbf{Data generating mechanism.} The final output of the data generating process consists $m$ matrices $\bM_1,\ldots,\bM_m$, where $\bM_i$ is a data of size $n_i\times d$, with $n_i$ being the numbers of observations in the $i$th cluster and $d$ is the number of variables. The first $d/2$ variables are set to be continuous and the remaining $d/2$ variables are binary. Each matrix $\bM_i$ is generated from a matrix of continuous data $\bM_i^{(c)}$ whose first $d/2$ columns are the same as the first $d/2$ columns of $\bM_i$, and the remaining $d/2$ columns are truncated to binary values corresponding to the last $d/2$ columns of $\bM_i$. Within a fixed cluster, the rows of $\bM_i$ are independently generated from a $d$-dimensional multivariate normal distribution with mean $\bmu_i$ and covariance matrix $\bSigma$. 

We now provide the details of the data generation setup. We set the number of clusters $m = 10$, let the numbers of observations in each clusters $n_1,\ldots,n_m$ be generated from $\mathrm{Binomial}(20, 1/2) + 100$, and the number of variables $d = 10$. Let $\bmu_1,\ldots,\bmu_m$ be generated independently from $\mathrm{N}_d(\zero_d, \eye_d)$ and for each $i \in\{1,\ldots,m\}$, and then we generate $m$ matrices $\bM_1^{(c)},\ldots,\bM_m^{(c)}$ as follows: The rows of $\bM_i^{(c)}$ are generated independently from the multivariate normal distribution $\mathrm{N}_d(\bmu_i, \bSigma)$, where $\bSigma = [\sigma_{kl}]_{k,l = 1,\ldots,d}$ is a sparse bandit matrix such that
\begin{align*}
&\sigma_{kk} = 5\quad\text{for }k = 1,\ldots,d,\\
&\sigma_{(k + 2)k} = \sigma_{k(k + 2)} = -1,\quad\text{for } k = 1,\ldots,d - 2,\\
&\sigma_{(k + 4)k} = \sigma_{k(k + 4)} = 1/2,\quad\text{for } k = 1,\ldots,d - 4,\\ 
&\sigma_{(k + 6)k} = \sigma_{k(k + 6)} = 1,\quad\text{for } k = 1,\ldots, d - 6.
\end{align*}
The resulting precision matrix $\bSigma^{-1}$ is also sparse, thereby introducing the conditional independence of these variables. More specifically, if the $(k, l)$th element of $\bSigma^{-1}$ is zero, then the $k$th variable and the $l$th variable are conditionally independent given the remaining variables. Namely, when one regresses the $k$th variable with respect to the remaining variables, it is desirable that a variable selection scheme is implemented such that the $l$th variable is ``filtered'' out. This feature is particularly attractive in the context of sequential hierarchical regression imputation with variable selection routines. In addition, we 
let $[\bM_i^{(c)}]_{jk}$ be the $(j, k)$th entry of $\bM_i^{(c)}$, $i=1,\ldots,m$, $j = 1,\ldots,n_i$, and $k = 1,\ldots,d$. Then we set $\bM_i$ as follows:
\begin{align*}
[\bM_i]_{jk} = \left\{
\begin{aligned}
&[\bM_i^{(c)}]_{jk},&\quad&\mbox{if }k \leq d / 2,\\
&1,&\quad&\mbox{if }k > d/2\mbox{ and }[\bM_i^{(c)}]_{jk} > 0,\\
&0,&\quad&\mbox{if }k > d/2\mbox{ and }[\bM_i^{(c)}]_{jk} < 0,
\end{aligned}
\right.
\end{align*} 
where $i = 1,\ldots,m$, $j = 1,\ldots,n_i$, and $k = 1,\ldots,d$. Namely, the first $d/2$ variables of $\bM_i$'s are taken directly from the first $d/2$ variables of $\bM_i^{(c)}$, and we convert the remaining $d/2$ variables to either $1$'s or $0$'s depending on whether their corresponding entries in $\bM_i^{(c)}$ are positive or negative. Consequently, $(\bM_i)_{i = 1}^m$ contains $d/2$ continuous variables and $d/2$ binary variables. The basic idea of this simulation setup is to consider multivariate missing data where the full conditional distribution of each variable given the remaining variable can be modeled directly or indirectly as a linear mixed-effect model and resembles the behavior of the sequential imputation model introduced earlier.

\textbf{Imposing missing values under MAR mechanism.} We generate the missing indicators sequentially as follows. For the first variable, we set 
\[
\prob([\bM_i]_{j1}\mbox{ is NA}) = 0.1,\quad 
i = 1,\ldots,m,\quad
j = 1,\ldots,n_i.
\] 
Then, for any $k = 2,\ldots,d$, the missing probability of $[\bM_i]_{jk}$ is given by
\[
\prob\left([\bM_i]_{jk}\mbox{ is NA}\mid[\bM_i]_{j(k - 1)}\right\} = \left\{
\begin{aligned}
&0,&\quad&\mbox{if }[\bM_i]_{j(k - 1)}\mbox{ is NA},\\
&\frac{1}{1 + \exp(-\alpha_{\mathrm{mis}} - \beta_{\mathrm{mis}}[\bM_i]_{j(k - 1)})},&\quad&\mbox{if }[\bM_i]_{jk}\mbox{ is observed.}
\end{aligned}
\right.
\]
Here, we set $\alpha_{\mathrm{mis}} = -3$ and $\beta_{\mathrm{mis}} = 1$ such that the overall missing percentages of $(\bM_i)_{i = 1}^n$ are roughly $10\%$. 

\textbf{Hypothetical analyst's model.} We posit the following hypothetical analyst's model that uses the last variable in $(\bM_i)_{i = 1}^m$ as the response variable and the first $(d - 1)$ variables as the covariates through the following generalized linear mixed-effect model with the logit link function:
\[
\mathrm{logit}(\prob([\bM_i]_{jd} = 1)) = 
\log\bigg\{\frac{\prob([\bM_i]_{jd} = 1)}{1 - \prob([\bM_i]_{jd} = 1)}\bigg\} =  \beta_0 + \sum_{k = 1}^{d - 1}[\bM_i]_{jk}\beta_k + b_i + \eps_{ij},
\]
where $b_1,\ldots,b_n\iidsim\mathrm{N}(0, 1)$ and $\eps_{ij}\iidsim\mathrm{N}(0, 1)$ for all $i = 1,\ldots,m$ and $j = 1,\ldots,n_i$.  Note that here, the practitioners only have access to the incomplete data. The overall goal is to investigate the performance of our proposed imputation method by inspecting the quality of the post-imputation estimation and inference procedures for $\bbeta$. 

We implement the SHRIMP method with spike-and-slab prior with $M = 10$ copies of imputed $(\bM_i)_{i = 1}^m$'s. For comparison, we also implement the \texttt{mice} package \citep{buuren2010mice} as well as the SuperMICE algorithm \citep{laqueur2022supermice} with the same number of imputed copies. Here, the SuperMICE algorithm is a sequential imputation method that uses ensemble learning algorithms to generate imputed values based on the predictive means and variances. For each imputed copy of $(\bM_i)_{i = 1}^m$, we estimate the fixed-effect regression coefficient $\bbeta$ using the \texttt{lme4} package \citep{lme4package} and draw inferences based on Rubin's combined rules \citep{rubinmultiple}. 
The entire numerical experiment is repeated for $100$ Monte Carlo replicates, and we compute the root-mean-squared errors (RMSEs), average standard errors (SEs), the empirical coverage rates (CRs), and the fractions of missing information (FMIs) for $\bbeta$. The true values of $\bbeta$ are obtained by averaging the estimates computed with the before-deletion data through the \texttt{lme4} package across repeated experiments. The results are tabulated in Table \ref{tab:ss_shrimp_simulation} below, which also includes the complete-case only analysis (CC only) for reference. We observe that the post-imputation estimation and inferential quality of the proposed method are comparable and sometimes outperform the baseline \texttt{mice} and SuperMICE methods together with the method using the complete cases of the data in terms of RMSEs, SEs, CRs, and FMIs. 

\begin{landscape}
\begin{table}[htbp]
\centering
\caption{Post-imputation estimation and inference for $\bbeta$ for Section \ref{sec:simulated_examples}: PB, RMSE, SE, CR, and FMI.}
\scalebox{0.8}{
\begin{tabular}{lc  |cccc  |ccccc  |ccccc  |ccccc}
\hline\hline
       \multicolumn{2}{c|}{$\bbeta$} &  \multicolumn{4}{c|}{CC only}  &  \multicolumn{5}{c|}{SS-SHRIMP} &  \multicolumn{5}{c|}{MICE}& \multicolumn{5}{c}{SuperMice} \\
\hline
       $\beta_k$&True value& PB&RMSE& SE & CR & PB&RMSE& SE & CR &FMI& PB&RMSE& SE & CR &FMI& PB&RMSE& SE & CR &FMI \\
\hline
 $\beta_0$&0.13& 14.1&0.77& 0.70& 0.94& 8.4&0.29& 0.34& 0.99&0.16&  1.8&0.35& 0.41 & 0.98& 0.18& 14.3&0.32&0.34&0.97&0.14\\
 $\beta_{1}$   &-0.006& 297.7&0.14& 0.14& 0.95& 32.5&0.07& 0.06& 0.92&0.13&  7.5&0.07& 0.08 & 0.91& 0.23& 29.6&0.08&0.06&0.87&0.10\\
 $\beta_{2}$   &0.05& 19.8&0.16& 0.14& 0.93& 17.1&0.07& 0.06& 0.95&0.15 &  18.9&0.06& 0.08 & 0.95& 0.26& 18.2&0.08&0.06&0.91&0.10\\
 $\beta_{3}$   &0.015& 59.6&0.16& 0.14& 0.93& 14.3&0.07& 0.06& 0.94&0.16&  0.6&0.07& 0.08 & 0.93& 0.26& 64.9&0.08&0.07&0.86&0.10\\
 $\beta_{4}$   &0.17& 17.5&0.17& 0.15& 0.94& 3.3&0.08& 0.06& 0.88&0.18&  0.2&0.07 & 0.08 & 0.93& 0.27& 11.4&0.08&0.06&0.93&0.10\\
 $\beta_{5}$   &0.06& 19.2&0.14& 0.14& 0.97& 2.4&0.06& 0.06& 0.97&0.14&  4.7&0.06& 0.08 & 0.98& 0.25& 8.0&0.08&0.06&0.90&0.10\\ 
 $\beta_{6}$   &0.31& 48.5&0.64& 0.58& 0.93& 13.3&0.23& 0.29& 0.98 &0.22&  4.3&0.27& 0.36 & 0.97& 0.25& 5.4&0.33&0.29&0.93&0.20\\
 $\beta_{7}$   &0.013& 398.8&0.66& 0.59& 0.95& 11.2&0.24& 0.29& 0.98&0.17&  9.4&0.29& 0.36 & 0.97& 0.19& 39.3&0.29&0.29&0.94&0.16\\
 $\beta_{8}$   &0.51& 15.5&0.59& 0.59& 0.98 & 8.5&0.28& 0.29& 0.95&0.16&  3.5&0.30& 0.36 & 0.94& 0.18& 0.9&0.27&0.29&0.97&0.15\\
 $\beta_{9}$  &0.09& 65.8&0.64& 0.59& 0.95& 14.4&0.25& 0.29& 0.98&0.17&  3.6&0.29& 0.37 & 0.95& 0.20& 33.2&0.28&0.29&0.93&0.17\\
\hline\hline
\end{tabular}
}
\label{tab:ss_shrimp_simulation}
\end{table}
\end{landscape}

\section{Application
	\label{sec:application}
} 
In this section, we apply the SHRIMP with spike-and-slab variable selection discussed in Section \ref{sec:SS_SHRIMP} to a real-world National Survey of Children's Health (NSCH) dataset. The dataset was taken from the National Survey of Children with Special Health Care Needs (NSCSHCN) in 2020. The dataset includes demographic variables such as gender, age, race, education level, poverty level, and insurance type, among others. The national survey data can be partitioned into different states (including the District of Columbia). We use the state variable as the cluster indicator so that the survey from each state forms one cluster. The number of observations in each state is approximately 750, with $m = 51$ clusters. Consequently, the entire NSCH dataset can be organized as a clustered data structure according to the problem formulation of the SHRIMP method. Because of the nature of survey data, most variables in this dataset also include missing values.
The overall goal of this section is to apply the MI method introduced in Section \ref{sec:SS_SHRIMP} to generate multiple copies of the ``completed'' NSCH dataset and draw MI-based subsequent inference. 

We take 9 variables that are considered significant among the entire surveyed variables that are quite relevant to the self-reported CSHCN variable (the extent to which the surveyed children are in severe health condition), convert the categorical variables into binary variables, and run the SHRIMP with spike-and-slab variable selection introduced in Section \ref{sec:SS_SHRIMP} with $10$ iterations to generate one copy of an imputed dataset. For comparison, we also implement the \texttt{mice} package, the SuperMICE algorithm, and the CC only analysis. For each of the proposed MI algorithm, the \texttt{mice} method, and the SuperMICE algorithm, we generate $M = 10$ copies of imputed datasets. Therefore, we obtain $M = 10$ copies of ``completed'' datasets as our MI outcomes. Regarding the selected variables, we follow \cite{https://doi.org/10.1002/sim.4355} and pick of demographical variables, including sex, age, race, Hispanic ethnicity, mother's education level, and the child's insurance type. The second group corresponds to the questionnaire items, where we focus on the partnership in the decision-making process, whether the surveyed child will receive comprehensive care within a medical home, and the maturity of the neighborhood amenities. Together with the CSHCN variable, the number of total variables under consideration is $d = 10$. 

Based on the MI copies generated above, we consider regressing the CSHCN variable against the other $9$ variables via a logistic mixed-effect model as the hypothetical analyst's model. Given an imputed copy of the dataset, we run the \texttt{lme4} package \citep{lme4package} to obtain the point estimates and the standard errors for the fixed-effect regression coefficients and apply Rubin's rule \citep{rubinmultiple} to draw the combined inference. We compare the results with the complete-case-only (cc'only) analysis results in Table \ref{tab:NSCH_SS_Shrimp}, where the metrics of interests include the point estimates of the regression coefficients, the corresponding $95\%$ confidence intervals (CIs), and the fraction of missing information (FMI) based on the SS-Shrimp MI method. The missing rate of each variable is in Table \ref{tab:NSCH_missing_rate}. 
\begin{landscape}
\begin{table}[htbp]
	\centering	
    \caption{Comparison between the MI analysis and the complete-case-only analysis for the regression coefficients of the selected variables (in percentages) }
	\scalebox{0.8}{
    \begin{tabular}{c|c|cccc|cccc|cccc|ccc}
		\hline\hline
        &&\multicolumn{4}{c|}{SS-SHRIMP} & \multicolumn{4}{c|}{MICE}&\multicolumn{4}{c|}{SuperMice}&
        \multicolumn{3}{c}{CC only}\\
		&Missing Rate(\%)&Est.&SE&P-value&FMI&Est.&SE&P-value&FMI&Est.&SE&P-value&FMI&Est.&SE&P-value\\
		\hline
        Intercept&&-1.70&0.08&0.00&1.9&-1.70&0.08&0.00&2.1&-1.65&0.08&0.00&3.1&-1.65&0.09&0.00\\
        Sex (Male)&0.08&0.30&0.03&0.00&0.4&0.30&0.03&0.00&0.2&0.31&0.03&0.00&0.3&0.34&0.03&0.00\\
        Age(0-5)  &0.00&-1.22&0.03&0.00&0.1&-1.22&0.03&0.00&0.2&-1.22&0.03&0.00&0.2&-1.24&0.04& 0.00\\
        Guardian Education&0.00&&&&&&&&&&&&&&&\\
        (College or higher)&&-0.24&0.03&0.00&0.3&-0.24&0.03&0.00&0.3&-0.24&0.03&0.00&0.1&-0.24&0.03&0.00\\
        Race (White) &0.47&-0.05&0.03&0.01&2.0&-0.05&0.03&0.1&0.5&-0.07&0.03&0.02&0.5&-0.07&0.04& 0.04\\
        Hispanic&0.36&-0.07&0.04&0.05&0.3&-0.08&0.04&0.05&1.0&-0.04&0.04&0.31&0.3&-0.05&0.04&0.24\\
        Insurance&1.49&&&&&&&&&&&&&&&\\
        Insured&&0.37&0.06&0.00&1.8&0.37&0.06&0.00&1.7&0.36&0.07&0.00&4.3&0.33&0.07&0.00\\
        Family partnered in&&&&&&&&&&&&&&&&\\
        decision making&0.92&1.72&0.03&0.00&1.6&1.72&0.03&0.00&1.7&1.71&0.03&0.00&0.7&1.74&0.03&0.00\\
        Child receives care&&&&&&&&&&&&&&&&\\
        in medical home&0.13&-0.25&0.03&0.00&0.1&-0.25&0.03&0.00&0.3&-0.27&0.03&0.00&0.4&-0.27&0.03&0.00\\
        Neighborhood &&&&&&&&&&&&&&&&\\
        amenities&3.13&0.03&0.04&0.47&7.1&0.04&0.04&0.33&6.0&0.02&0.04&0.63&3.3&0.03&0.05&0.52\\
		\hline\hline
	\end{tabular}
    }
	\label{tab:NSCH_SS_Shrimp}
\end{table}
\end{landscape}

\section{Discussion
\label{sec:discussion}
} 

We have illustrated that the variable selection problem in the presence of missing response variables in mixed-effects regression models can be done by a hierarchical Bayesian approach with a spike-and-slab prior distribution for the linear coefficients. We successfully derive an efficient Gibbs sampler for posterior computation of the corresponding linear and logistic mixed-effects models. The hierarchical Bayesian model itself also permits the integration with the sequential hierarchical regression imputation strategy introduced by \cite{yucel2017sequential} for multiple imputations of the missing responses, further facilitating the computational efficiency of the corresponding MCMC algorithm. 

There are some potential future extensions of the current methodology. The numerical examples provided in this work are relatively low-dimensional regression problems. Although the spike-and-slab prior distributions \eqref{eqn:spike_and_slab_prior} permits the derivation of closed-form Gibbs sampler either by a direct approach or via a PX-DA strategy (e.g., the auxiliary P\'olya-Gamma random variable), the corresponding computation expense for the MCMC is still problematic with ultra-high-dimensional data. Even with the help of Monte Carlo sampling methods and the spike-and-slab prior \eqref{eqn:spike_and_slab_prior}, it is still required to explore the entire space of all possible models as much as possible. Nonetheless, the complexity of the space of all possible models grows exponentially with the number of predictors, and in moderately high-dimensional setups, the MCMC could be cumbersome or even infeasible to implement. 
We have already observed the potential computational difficulty of the MCMC-based MI method involving variable selection in the simulated examples. 
In particular, we note that the computation expense for the spike-and-slab variable selection composite with SHRIMP for MI is much more expensive than the other competitors, but we gain estimation and variable selection accuracy instead.
It has also been pointed out in \cite{castillo2015bayesian} that algorithms that can successfully address ultra-high-dimensional variable selection problems are beyond the scope of fully Bayesian methods. 

In contrast to relying on MCMC-based posterior computation algorithms, which is a class of exact Bayesian inference methods in the sense that the random samples drawn from the Markov chain can be regarded as samples generated from the exact full posterior distribution, a relatively more efficient method is the variational inference (VI). Unlike the MCMC approach, the VI is an approximate Bayesian inference algorithm that can be much faster but at the cost of certain model bias. Under certain regularity conditions, it has also been proved that the variational posterior distribution is comparable to the exact posterior distribution \citep{zhang2020convergence,pati2018statistical,wang2019frequentist,han2019statistical}. The use of VI for linear regression models has been restricted in the case of low-dimensional models \citep{you2014variational}. In the future, we plan to explore the methodology and theory for VI for linear and generalized linear mixed-effects models in the presence of the missing responses for the sake of computational efficiency for high-dimensional data. 


\section*{Acknowledgements}
We want to thank\ldots

\section*{Declaration of conflicting interests}
The authors declared no potential conflicts of interest with respect to the research, authorship and/or
publication of this article.

\section*{Funding}
This is the place to mention the funding if it is applicable for the paper.

\appendix
\section*{Appendix}
\section{Gibbs sampler for Section \ref{sec:linear_mixed_effects_regression_models}} 
\label{sec:gibbs_sampler_for_section_sec:linear_mixed_effects_regression_models}
In this section, we derive the detailed Gibbs sampling algorithm, which reduces to the following full conditional distributions of the parameters $\btheta = (\bbeta, \sigma_b, \sigma_e)$ and the random effects $\bb = [b_1,\ldots,b_m]\transpose$. A single cycle of the Gibbs sampler iterates the following sampling schemes:
\begin{align}
\nonumber
&(\beta_k\mid\bX, \bbeta_{-k}, \bb, \sigma_b,\sigma_e, w, \mu_0, \sigma_0)\\
&\quad\sim
\left\{
\begin{array}{ll}
 w^*_1\delta_0 + w^*_2\mathrm{N}(\widehat\mu, \widehat V),&\text{if the }k\text{th variable is undetermined},\\
 \mathrm{N}(\widehat\mu, \widehat V),&\text{if the }k\text{th variable is forced to be selected},
\end{array}
\right.
\end{align}
\begin{align}
\label{eqn:Gibbs_w}
&(w\mid \bX, \bbeta)\sim\mathrm{Beta}\left(a_w + \sum_{k = 1}^pz_k, b_w + \sum_{k = 1}^p(1 - z_k)\right),\\
\label{eqn:Gibbs_mu0}
&(\mu_0\mid \bbeta, \sigma_0)\sim\mathrm{N}\left(\left(1 + \frac{1}{\sigma_0^2}\sum_{k = 1}^pz_k\right)^{-1}\frac{1}{\sigma_0^2}\sum_{k = 1}^p\beta_k, 
\left(1 + \frac{1}{\sigma_0^2}\sum_{k = 1}^pz_k\right)^{-1}
\right),\\
\label{eqn:Gibbs_sigma0}
&(\sigma_0^2\mid \bbeta, \mu_0)\sim\mbox{Inverse-Gamma}\left(1 + \frac{1}{2}\sum_{k = 1}^pz_k, 1 + \frac{1}{2}\sum_{k = 1}^pz_k(\beta_k - \mu_0)^2\right),\\
\label{eqn:Gibbs_bi}
&(b_i\mid \bX, \bbeta, \sigma_b, \sigma_e)\sim \mathrm{N}(\widehat b_i, V(\widehat b_i)),\\
\label{eqn:Gibbs_sigmae}
&(\sigma_e^2\mid \bX, b_1, \ldots, b_m, \bbeta, \sigma_b)\sim \left(1 + \sum_{i = 1}^m\sum_{j =1}^n\widehat\eps_{ij}^2\right)\chi_{\nu_e + mn - 1}^{-2},\\
\label{eqn:Gibbs_sigmab}
&(\sigma_b^2\mid \bX, b_1, \ldots, b_m, \bbeta, \sigma_e)\sim \left(\frac{\nu_b + \sum_{i = 1}^mb_i^2}{\nu_b + m}\right)\chi_{\nu_b + m}^{-2},
\end{align}
where $\bX$ denotes the full set of covariates $\bX = [\bx_{ij}]_{i = 1,\ldots,m, j = 1,\ldots,n}$, $z_k = \mathbbm{1}(\beta_k \neq 0)$, $w_1^*, w_2^*, \widehat{V}, \widehat{\mu}$ are the same as those given in Section \ref{sec:linear_mixed_effects_regression_models},
\begin{align*}
&\widehat\eps_{ij} = y_{ij} - \bx_{ij}\transpose \widehat\bbeta, \quad\widehat\bbeta = \left(\sum_{i = 1}^m\sum_{j = 1}^n\bx_{ij}\bx_{ij}\transpose\right)^{-1}\sum_{i = 1}^m\sum_{j = 1}^n\bx_{ij}(y_{ij} - b_i), \\
&V(\widehat b_i) = \left(\frac{n}{\sigma_e^2} + \frac{1}{\sigma_b^2}\right)^{-1},
\quad
\widehat b_i = \frac{V(\widehat b_i)}{\sigma_e^2}\sum_{j = 1}^n(y_{ij} - \bx_{ij}\transpose\bbeta).
\end{align*}
Note that formulas \eqref{eqn:Gibbs_bi}, \eqref{eqn:Gibbs_sigmae}, and \eqref{eqn:Gibbs_sigmab} are the same as those appearing in Section 2.2.1 in \cite{yucel2017sequential}. The last step in one iteration of the Gibbs sampler is to draw the predictive posterior distribution of the missing response $y_{ij}\in (y_{\mathrm{mis}})$ using the SHRIMP strategy described at the end of Section \ref{sec:linear_mixed_effects_regression_models}. 


\section{Gibbs sampler for Section \ref{sec:logistic_mixed_effects_regression_models}} 
\label{sec:gibbs_sampler_for_section_sec:logistic_mixed_effects_regression_models}

We provide the complete full conditional distributions that are required for the Gibbs sampler to draw posterior samples from the joint distribution of $(\bbeta, b_1,\ldots,b_m)$, together with the samples of the missing data $(y_{\mathrm{mis}})$. Following the derivation in Section \ref{sec:logistic_mixed_effects_regression_models},
we obtain the following closed-form full conditional distribution of $\bbeta$, $w$, $\mu_0$, and $\sigma_0^2$:
\begin{align}
\nonumber
&(\beta_k\mid\bX, \bY, \bOmega, \bbeta_{-k}, \bb,\sigma_b, w, \mu_0, \sigma_0)\\
&\quad\sim
\left\{
\begin{array}{ll}
 w^*_1\delta_0 + w^*_2\mathrm{N}(\widehat\mu, \widehat V),&\text{if the }k\text{th variable is undetermined},\\
 \mathrm{N}(\widehat\mu, \widehat V),&\text{if the }k\text{th variable is forced to be selected},
 \end{array}
\right.
\\
\label{eqn:binary_Gibbs_w}
&(w\mid \bX, \bbeta)\sim\mathrm{Beta}\left(a_w + \sum_{k = 1}^pz_k, b_w + \sum_{k = 1}^p(1 - z_k)\right),\\
\label{eqn:binary_Gibbs_mu0}
&(\mu_0\mid \bbeta, \sigma_0)\sim\mathrm{N}\left(\left(1 + \frac{1}{\sigma_0^2}\sum_{k = 1}^pz_k\right)^{-1}\frac{1}{\sigma_0^2}\sum_{k = 1}^p\beta_k, 
\left(1 + \frac{1}{\sigma_0^2}\sum_{k = 1}^pz_k\right)^{-1}
\right),\\
\label{eqn:binary_Gibbs_sigma0}
&(\sigma_0^2\mid \bbeta, \mu_0)\sim\mbox{Inverse-Gamma}\left(1 + \frac{1}{2}\sum_{k = 1}^pz_k, 1 + \frac{1}{2}\sum_{k = 1}^pz_k(\beta_k - \mu_0)^2\right)
\end{align}
where $\bX$ denotes the full set of covariates $\bX = [\bx_{ij}]_{i = 1,\ldots,m, j = 1,\ldots,n}$, $z_k = \mathbbm{1}(\beta_k \neq 0)$, and
 the formulas for computing $w_1^*, w_2^*, \widehat{V}, \widehat{\mu}$ are provided in Section \ref{sec:logistic_mixed_effects_regression_models}. 
The full conditional distribution of the auxiliary variables $\bOmega = [\omega_{ij}]_{m\times n}$ is given by 
\eqref{eqn:binary_Gibbs_omega} in Section \ref{sec:logistic_mixed_effects_regression_models}.
Similar to the derivation of \eqref{eqn:binary_Gibbs_beta}, the full conditional distributions of the random effects $b_1,\ldots,b_m$ can be derived analogously:
\begin{align*}
p(b_i\mid \bX, \bY, \bbeta, \sigma_b)& \propto p(b_i)\prod_{j = 1}^n\calL(\eta_{ij}\mid y_{ij})\\
& \propto p(b_i)\prod_{j = 1}^n\exp\left\{
\left(y_{ij} - \frac{1}{2}\right)(\bx_{ij}\transpose\bbeta + b_i)-\frac{\omega_{ij}}{2}(\bx_{ij}\transpose\bbeta + b_i)^2\right\}\\
& \propto p(b_i)\prod_{j = 1}^n\exp\left\{
-\frac{\omega_{ij}}{2}\left[b_i^2 -2\left(\frac{y_{ij} - 1/2}{\omega_{ij}} - \bx_{ij}\transpose\bbeta\right)b_i\right]\right\}\\
& \propto p(b_i)\prod_{j = 1}^n\exp\left[
-\frac{\omega_{ij}}{2}\left(b_i - u_{ij}\right)^2\right],
\end{align*}
where $u_{ij} = (y_{ij} - 1/2)/\omega_{ij} - \bx_{ij}\transpose\bbeta$. Since $p(b_i) = (1/\sqrt{2\pi\sigma_b^2})\exp[-b_i^2/(2\sigma_b^2)]$, it follows directly from the normal conjugacy that
\begin{align}
(b_i\mid\bX, \bY, \bbeta, \sigma_b)\sim \mathrm{N}\left(\left(\frac{1}{\sigma_b^2} + \sum_{j = 1}^n\omega_{ij}\right)^{-1}\sum_{j = 1}^n\omega_{ij}u_{ij}, \left(\frac{1}{\sigma_b^2} + \sum_{j = 1}^n\omega_{ij}\right)^{-1}\right).
\end{align}
The full conditional distribution of $\sigma_b$ is the same as \eqref{eqn:Gibbs_sigmab}:
\[
(\sigma_b^2\mid \bX, b_1, \ldots, b_m)\sim \left(\frac{\nu_b + \sum_{i = 1}^mb_i^2}{\nu_b + m}\right)\chi_{\nu_b + m}^{-2}.
\]
 The last step in a single iteration of the Gibbs sampler is to draw the predictive posterior distribution of the missing response $y_{ij}\in (y_{\mathrm{mis}})$ following the SHRIMP strategy mentioned at the end of Section \ref{sec:linear_mixed_effects_regression_models}. 


\bibliography{reference}

\end{document}